\documentclass[prd,aps,twocolumn,preprintnumbers,amsmath,amssymb,superscriptaddress]{revtex4}
\pdfoutput=1
\usepackage{amsmath}
\usepackage{graphicx}
\usepackage{dcolumn}
\usepackage{bm}
\usepackage{amssymb}
\usepackage{latexsym}
\usepackage{epstopdf}
\usepackage{threeparttable}
\usepackage{booktabs}
\usepackage{tabularx}
\usepackage{mathrsfs}
\usepackage{hhline}
\usepackage{multirow}
\usepackage{color}
\usepackage{lipsum}
\usepackage[colorlinks,linkcolor=magenta,anchorcolor=blue,citecolor=blue]{hyperref}
\usepackage{ulem}

\def\be{\begin{equation}}
\def\ee{\end{equation}}
\def\bea{\begin{eqnarray}}
\def\eea{\end{eqnarray}}

\begin{document}
\title{Adiabatic Duality: Duality of cosmological models with varying slow-roll parameter and sound speed}

\author{Jiaming Shi}
\email{2016jimshi@mails.ccnu.edu.cn}
\affiliation{Institute of Astrophysics, Central China Normal University, Wuhan 430079, China}

\author{Zheng Fang}
\email{ifz@mails.ccnu.edu.cn}
\affiliation{Institute of Astrophysics, Central China Normal University, Wuhan 430079, China}

\author{Taotao Qiu}

\email[Corresponding author: ]{qiutt@hust.edu.cn}
\affiliation{School of Physics, Huazhong University of Science and Technology, Wuhan, 430074, China}
\affiliation{Institute of Astrophysics, Central China Normal University, Wuhan 430079, China}

\begin{abstract}
There have been thousands of cosmological models for our early universe proposed in the literature, and many of them claimed to be able to give rise to scale-invariant power spectrum as was favored by the observational data. It is thus interesting to think about whether there are some relations among them, e.g., the duality relation. In this paper, we investigate duality relations between cosmological models in framework of  general relativity (GR) , not only with varying slow-roll parameter $\epsilon$, but also with sound speed $c_s$, which can then be understood as ``{\it adiabatic duality}". Several duality relationships are formulated analytically and verified numerically. We show that models with varying $\epsilon$ and constant $c_s$ can be dual in scalar spectral index, but not tensor one. On the other hand, allowing both $\epsilon$ and $c_s$ to vary can make models dual in both scalar and tensor spectral indices.
\end{abstract}

\maketitle

\section{introduction}
It is always interesting to ask what our universe has been like at its earliest stage. A most acceptable answer might be that it experienced a period of inflation era \cite{Guth:1980zm, Linde:1981mu, Starobinsky:1980te}, for it provides a solution to many Big-Bang problems. Nevertheless, due to its incapability of solving another issue of Big-Bang Scenario known as the singularity problem \cite{Hawking:1969sw, Borde:1993xh}, many other scenarios/models are coming out constantly as its complementary/alternative candidates, such as bounce \cite{Novello:2008ra}, ekpyrotic \cite{Khoury:2001wf}, slow expansion \cite{Piao:2003ty} and so on. These bring our early universe with full of possibilities.

For these models, what is among the most important things is the necessity of being consistent with the observational data. Especially, there has been precise evidence that the scalar perturbations are nearly scale-invariant, with only few percent level of deviation \cite{Akrami:2018odb}. Decades before, it had been thought that there were only two possibilities that could obtain nearly scale-invariance, namely slow-roll inflation and matter bounce (bounce with matter-like contracting phase before the universe's expansion) \cite{Finelli:2001sr, Wands:1998yp}, while ekpyrotic and slow expansion models suffered from blue power spectrum \cite{Lyth:2001pf, Piao:2004jg}. However, it has later been realized that, by requiring a varying slow-roll parameter $\epsilon$ \footnote{Strictly speaking, this terminology only applies for slow-roll inflation, for in other cases or when the slow-roll parameter is varying, the universe might not ``slow-roll" any longer. However, we still use the name to stand for the very parameter.} in these models, scale-invariance can be recovered again \cite{Khoury:2009my, Piao:2010bi, Joyce:2011ta}. The reason that varying $\epsilon$ can promote scale-invariance of the power spectrum is simply due to the fact that it can get involved in the perturbation action and change the behavior of the perturbation equations, which is also known as ``adiabatic mechanism" (\cite{Khoury:2009my}, see also \cite{Linde:2009mc, Khoury:2011ii} for a debate).  Further study shows that, for models constructed under GR, scale-invariance will be obtained as long as the condition:
\be
\frac{(a\sqrt{|\epsilon|})^{\prime\prime}}{a\sqrt{|\epsilon|}}=\frac{2}{|\tau|^2}~,
\ee
is satisfied, where $a$ is the scale factor and $\tau$ is the conformal time, $\tau\equiv \int a^{-1}(t)dt$. With the varying behavior of $\epsilon$, the constraint on $a$ by this condition gets loosen, making more cosmic evolutions allowable.

The adiabatic mechanism can also be applied to inflation model itself. Recently, there is a model attracting people's eyes called the ultra-slow-roll inflation \cite{Kinney:2005vj, Namjoo:2012aa}. It possesses an ``exact" flat potential, namely $dV/d\phi=0$, which furtherly results in a decreasing $\epsilon$, namely $\epsilon\sim a^{-6}$. Although as an inflation model it is not necessary to have varying $\epsilon$, this interestingly make the behavior of its perturbations like those of {\it matter contraction phase} in \cite{Finelli:2001sr, Wands:1998yp}, or {\it slow expansion phase} in \cite{Piao:2003ty},  which is dominated by its growing mode, rather than the constant one. It implies some links between inflation and other cosmological models.

Other than $\epsilon$, the behavior of power spectrum can also be affected by the sound speed $c_s$. The sound speed is a factor in front of the spatial derivative of the perturbations in the equation of motion, therefore, different from $\epsilon$ which modifies the background evolution, the sound speed modifies the effective horizon, as well as the values when perturbations exit and reenter the horizon. For perturbations that are not conserved outside the horizon, such a modification will correspondingly affect the scale-variance of the power spectrum. Therefore, if we furtherly allow $c_s$ to vary, we may have even more models with scale-invariant power spectrum \cite{ArmendarizPicon:2003ht, Piao:2006ja, Magueijo:2008pm, Khoury:2008wj, Lu:2009he}.

Although the current work focuses on the framework of GR, as a side remark, let's also mention that when the modified gravity is taken into account, the possibility of getting scale-invariance will also be enlarged, due to the fact that the scale factor can be corrected by a conformal factor $F$, namely $a\rightarrow a\sqrt{F}$. For relevant works, see \cite{Piao:2011mq, Qiu:2012ia, Ijjas:2015zma, Luan:2019xer}.

Given the more-than-enough models that can meet with the current observational data, as an extension, we would ask: can there be more links between those models? Especially, among the models with varying $\epsilon$ and $c_s$, will they have some relations such as dualities? Actually, there have been many papers discussing on dualities between early universe models, for example, \cite{Wands:1998yp} showed us the duality between slow-roll inflation and matter-like contraction, \cite{Piao:2004uq} discussed the dualities of the primordial perturbation spectra from various expanding/contracting phases with constant $\epsilon$, while \cite{Khoury:2010gw} presented the duality between ekpyrosis with varying $\epsilon$ is dual to inflation with constant $\epsilon$. In \cite{Morse:2018kda, Gao:2019sbz, Lin:2019fcz}, there are also debates on whether there is duality between slow-roll inflation and ultra-slow-roll inflation models. In this paper, we try to investigate as a whole the duality among varying $\epsilon$ and $c_s$ models, in order to see whether such nontrivial parameters will bring us anything new about the duality relations. Since these duality relations are based on the aforementioned ``adiabatic mechanism", they can be called as kind of ``adiabatic duality", in contrast to the ``conformal duality" studied in \cite{Piao:2011mq, Qiu:2012ia, Ijjas:2015zma, Luan:2019xer}. 

The rest of the paper is arranged as follows: in Sec. II we show the formulation of perturbations from a single field cosmological model in general case. In Sec III we focus on the duality for varying $\epsilon$ and constant $c_s$ models, while in Sec. IV we extend our discussion to the case where both $\epsilon$ and $c_s$ are varying. In Sec. V we check our analysis by performing numerical calculation of the perturbation equations. Sec. VI is the final conclusions and discussions.

\section{perturbations from a single field model}

We will consider the linear perturbations generated in the early universe, which is described by the action
\be
\label{action}
S=\frac{1}{2}\int d^4x\sqrt{-g}\left[M_p^2R+2P(\phi,X)\right]~
\ee
where $\phi$ is a scalar field while $X\equiv-\partial_\mu\phi\partial^\mu\phi/2$. Hereafter we choose the unit $M_p^2=1$.  As there is only one scalar degree of freedom in this kind of model, the scalar perturbations are purely adiabatic. A tedious but conventional calculation shows that such adiabatic perturbations obey the perturbation equation:
\be
\label{equation}
u^{\prime\prime}+\left(c_s^2k^{2}-\frac{z^{\prime\prime}}{z}\right)u=0~,
\ee
where we define the perturbation variable $u\equiv z\zeta$, with $\zeta$ denoting the curvature perturbation, and $z\equiv  a\sqrt{|\epsilon|}/c_s$. The slow-roll parameter $\epsilon$ is defined as $\epsilon\equiv-\dot H/H^2$, where $H$ is the Hubble parameter, and the sound speed squared $c_s^2$ is defined as
\be
c_s^2\equiv \frac{P_{,X}}{(2XP_{,X}-P)_{,X}}~.
\ee
Moreover, the prime in Eq. (\ref{equation}) means derivative with respect to conformal time $\tau$.

In usual case where $z(\tau)$ can be parameterized as a power-law form of $\tau$, one in general has $z^{\prime\prime}/z\propto|\tau|^{-2}$. Therefore, it is reasonable to set
\be
\label{nuz}
\frac{z^{\prime\prime}}{z}=\frac{4\nu_z^{2}-1}{4|\tau|^{2}}~,
\ee
where $\nu_z$ is a parameter. Moreover, we assume that the $c_s$ also has a power-law form of $\tau$, namely $c_s\sim (-\tau)^s$  with $s$ the power index, then Eq. (\ref{equation}) has the Hankel-function solution:
\be
u\simeq\sqrt{|\tau|}[c_{1}H_{\nu}(\int c_s kd\tau)+c_{2}H_{-\nu}(\int c_s kd\tau)]
\ee
where $\nu\equiv \nu_z/(s+1)$. Here $s+1>0$ is required in order to ensure that the fluctuation modes can exit the sound horizon. Note that in general the index of the Hankel function $\nu$ is different from $\nu_z$, however for constant $c_s$ case where $s=0$, the two indices coincide with each other. Moreover, comparing with the initial condition solution
\be
u_{ini}=\frac{1}{\sqrt{2c_s k}}e^{i\int c_s k d\tau}~
\ee
which is obtained from Eq. (\ref{equation}) in the $k\rightarrow \infty$ limit, one can fix the coefficients $c_1=c_2=\sqrt{\pi/(s+1)}/2$. Therefore the power spectrum can be obtained as:
\bea
P_{S}&\equiv&\frac{k^{3}}{2\pi^{2}}\Bigg|\frac{u}{a\sqrt{|\epsilon|}/c_s}\Bigg|^{2}~\nonumber\\
&\sim&\frac{(s+1)^2H_\ast^2}{8\pi^2|\epsilon_\ast| c_{s\ast}}\left(\frac{\tau}{\tau_\ast}\right)^{-(3-2\nu)(s+1)}|c_sk\tau|^{3-2|\nu|}~,
\eea
with the spectral index
\be
n_S\equiv \frac{d\ln P_S}{d\ln k}=3-2|\nu|~,
\ee
where ``$\ast$" means values taken at some pivot timepoint $\tau=\tau_\ast$. From the expression one can easily see that, both $\nu$ and $-\nu$ can give rise to the same spectral index. Moreover, to get the scale invariant power spectrum which is favored by the observational data, we need to have $|\nu|=3/2$. In the case where $\epsilon$ and $c_s$ are constants, this requires either  $a \sim (-\tau)^{-1}$ or $a \sim (-\tau)^{2}$ \cite{Wands:1998yp}. In the former case, the perturbations are dominated by their constant mode, which makes their behavior like those in slow-roll inflation regime, while in the latter case those are dominated by their growing model, like a matter-dominated contraction. However, as we will see below, for varying $\epsilon$ and $c_s$, case may be different.

We also consider the tensor perturbation generated by model (\ref{action}), which is important as it provides the primordial gravitational waves that we're searching for. The tensor perturbation equation can be derived from the action (\ref{action}) as
\be
\label{tensorEoM}
v^{\prime\prime}+\left(k^2-\frac{a^{\prime\prime}}{a}\right)v=0~,~\frac{a^{\prime\prime}}{a}=\frac{4\nu_T^2-1}{4|\tau|^2}~,
\ee
where $v\equiv ah/2$ and $h$ is the tensor mode of the metric perturbation. Note that since we restrict ourselves in the case of GR, the sound speed of tensor perturbation is unity. Similar calculation shows that the power spectrum for tensor perturbation is:
\bea
P_T&\equiv&\frac{k^{3}}{\pi^{2}}\Big|\frac{v}{a/2}\Big|^{2}~\nonumber\\
&\sim&\frac{2H_\ast^2}{\pi^2}\left(\frac{\tau}{\tau_\ast}\right)^{-(3-2\nu_T)}|k\tau|^{3-2|\nu_T|}~,
\eea
with the tensor spectral index
\be
n_T\equiv \frac{d\ln P_T}{d\ln k}=3-2|\nu_T|~.
\ee

In practical analysis and observations, people are used to express the tensor spectrum in terms of the tensor-scalar ratio, which is
\be
r\equiv \frac{P_T}{P_S}=\frac{16\epsilon_\ast |c_{s\ast}|}{(s+1)^2}\left(\frac{\tau}{\tau_\ast}\right)^{\eta+s}~.
\ee

\section{Cosmic duality for varying $\epsilon$ and constant $c_s$}

As a first step, we now consider the case where the slow-roll parameters are varying while the sound speed remains constant. From the very definition of the slow-roll parameter, one thus derives the expression of $\epsilon$ in terms of conformal time $\tau$ as:
\be
\label{epsilon}
\epsilon=1-\frac{\mathcal{H}^{\prime}}{\mathcal{H}^{2}}=1+\left(\frac{1}{\mathcal{H}}\right)^{\prime}~,
\ee
where $\mathcal{H}$ is conformal Hubble parameter, $\mathcal{H}\equiv aH$.
The conformal time $\tau$ will be negative, with its absolute value $|\tau|=-\tau$ decreasing. Assuming $\epsilon(\tau)=\epsilon_{0}(-\tau)^{\eta}$, one can solve (\ref{epsilon}) to get

\bea
\frac{1}{\mathcal{H}(\tau)}&\simeq&\frac{(-\tau)(1+\eta-\epsilon)}{1+\eta}~,\nonumber\\
\mathcal{H}(\tau)&\simeq&\frac{1+\eta}{(-\tau)(1+\eta-\epsilon)}~.
\eea

Here the approximation in computing $\mathcal{H}(\tau)$ is due to that the integration constant has been neglected. This approximation is acceptable as we will verify our final analytical result of spectral index by numerical simulations later.  Moreover, according to $\mathcal{H}\equiv a^\prime(\tau)/a(\tau)$, one also has

\be
\label{scalefactor}
a(\tau)=e^{\int\mathcal{H}d\tau}\simeq(-\tau)^{-1}|1+\eta-\epsilon|^{1/\eta}~.
\ee

Substituting the expressions of $a(\tau)$ and $\epsilon(\tau)$ into the expression of $z$ (where we set $c_s=1$) one has:
\bea
\label{z}
z&\simeq&\sqrt{|\epsilon_{0}|}(-\tau)^{\frac{\eta}{2}-1}|1+\eta-\epsilon_{0}(-\tau)^{\eta}|^{1/\eta}~, \nonumber\\
\frac{z^{\prime\prime}}{z}&=&\frac{1}{4}(1-\epsilon+\eta)^{-2}(-\tau)^{-2}[(1+\eta)^2(\eta-2)(\eta-4) \nonumber\\
&&-2\epsilon(2-\eta^2+\eta^3)+\epsilon^2\eta(\eta-2)]~,
\eea
and from (\ref{nuz}) we get
\bea
\label{nu}
\nu=\nu_z&=&\pm\frac{1}{2}\left(\frac{\epsilon-3(1+\eta)}{\epsilon-(1+\eta)}-\eta\right)\nonumber\\
&&\times\sqrt{1-\frac{4\epsilon\eta(\eta+1)}{[\epsilon(1-\eta)+(\eta+1)(\eta-3)]^{2}}}~\nonumber\\
&\simeq&\pm\frac{1}{2}\left(\frac{\epsilon-3(1+\eta)}{\epsilon-(1+\eta)}-\eta\right)~.
\eea
In deriving the second step, notice that we're using the 0-th order approximation for varying $\epsilon$, where the last term in square root can be ignored safely either the approximate value becomes too large or too small. Note also that when $\eta=0$, $\epsilon$ becomes constant and the result recovers the usual one of $2\nu=\pm(\epsilon-3)/(\epsilon-1)$.

Ref. \cite{Wands:1998yp} has pointed out that any two scenarios giving opposite $\nu$ will become dual to each other, for they give rise to the same power spectrum. Here, we revisit this remark for scenarios with varying $\epsilon$. For two scenarios with
\be
\nu=\frac{1}{2}\left(\frac{\epsilon-3(1+\eta)}{\epsilon-(1+\eta)}-\eta\right)~,~~~\tilde{\nu}=\frac{1}{2}\left(\frac{\tilde{\epsilon}-3(1+\tilde{\eta})}{\tilde{\epsilon}-(1+\tilde{\eta})}-\tilde{\eta}\right)~,
\ee
a dual relation between to two is $|\nu|=|\tilde{\nu}|$, namely
\be
\label{dualscalar}
\frac{\epsilon-3(1+\eta)}{\epsilon-(1+\eta)}-\eta=\pm\left(\frac{\tilde{\epsilon}-3(1+\tilde{\eta})}{\tilde{\epsilon}-(1+\tilde{\eta})}-\tilde{\eta}\right)~.
\ee

We first consider the case where $``\pm"\rightarrow``-"$ in Eq. (\ref{dualscalar}). Since now both $\epsilon$ and $\tilde{\epsilon}$ are varying, an interesting case is that they approach to different directions. For $\epsilon(\tau)\rightarrow \pm\infty$ and $\tilde{\epsilon}(\tilde{\tau})\rightarrow0$, or vice versa, one has:
\be
\label{dual1}
\eta+\tilde{\eta}=4~.
\ee
which is a duality relation between $\eta$ and $\tilde{\eta}$. Considering the constraint of scale invariance of the power spectrum, namely $3-2|\nu|=0$, we have the following possibilities:\\
$\bullet$ $\nu=-\tilde{\nu}=3/2$, which leads to $\eta=4$, $\tilde\eta=0$. \\
$\bullet$ $\nu=-\tilde{\nu}=-3/2$, which leads to $\eta=-2$, $\tilde\eta=6$. \\
There are also non-trivial possibilities for $\epsilon$ and $\tilde{\epsilon}$ approaching to the same direction. For example, for both $\epsilon$ and $\tilde\epsilon$ approaching to $\pm\infty$, one has:
\be
\label{dual2}
\eta+\tilde\eta=2~,
\ee
and considering the constraint of scale invariance, we have $\eta=-2$ and $\tilde\eta=4$.
For both $\epsilon$ and $\tilde\epsilon$ approaching to $0$, one has:
\be
\label{dual3}
\eta+\tilde\eta=6~,
\ee
and considering the constraint of scale invariance, we have $\eta=0$ and $\tilde\eta=6$.

Another duality relation arises for $``\pm"\rightarrow``+"$ in Eq. (\ref{dualscalar}).
Note that this becomes trivial for constant $\epsilon$ and will give $\epsilon=\tilde{\epsilon}$ only. However, for varying $\epsilon$, by requiring $\epsilon$ and $\tilde{\epsilon}$ approaching to different directions ($\epsilon(\tau)\rightarrow \pm\infty$ and $\tilde{\epsilon}(\tilde{\tau})\rightarrow0$, or vice versa) one has:
\be
\label{dual4}
|\eta-\tilde{\eta}|=2~.
\ee
Considering the scale invariance, we have the following possibilities:\\
$\bullet$ $\nu=\tilde\nu=3/2$, which leads to $\eta=4$, $\tilde\eta=6$. \\
$\bullet$ $\nu=\tilde\nu=-3/2$, which leads to $\eta=-2$, $\tilde\eta=0$. \\
Moreover, if $\epsilon$ and $\tilde{\epsilon}$ approach to the same direction, it gives trivial result as well.

\begin{figure}
\centering
\includegraphics[scale=0.4]{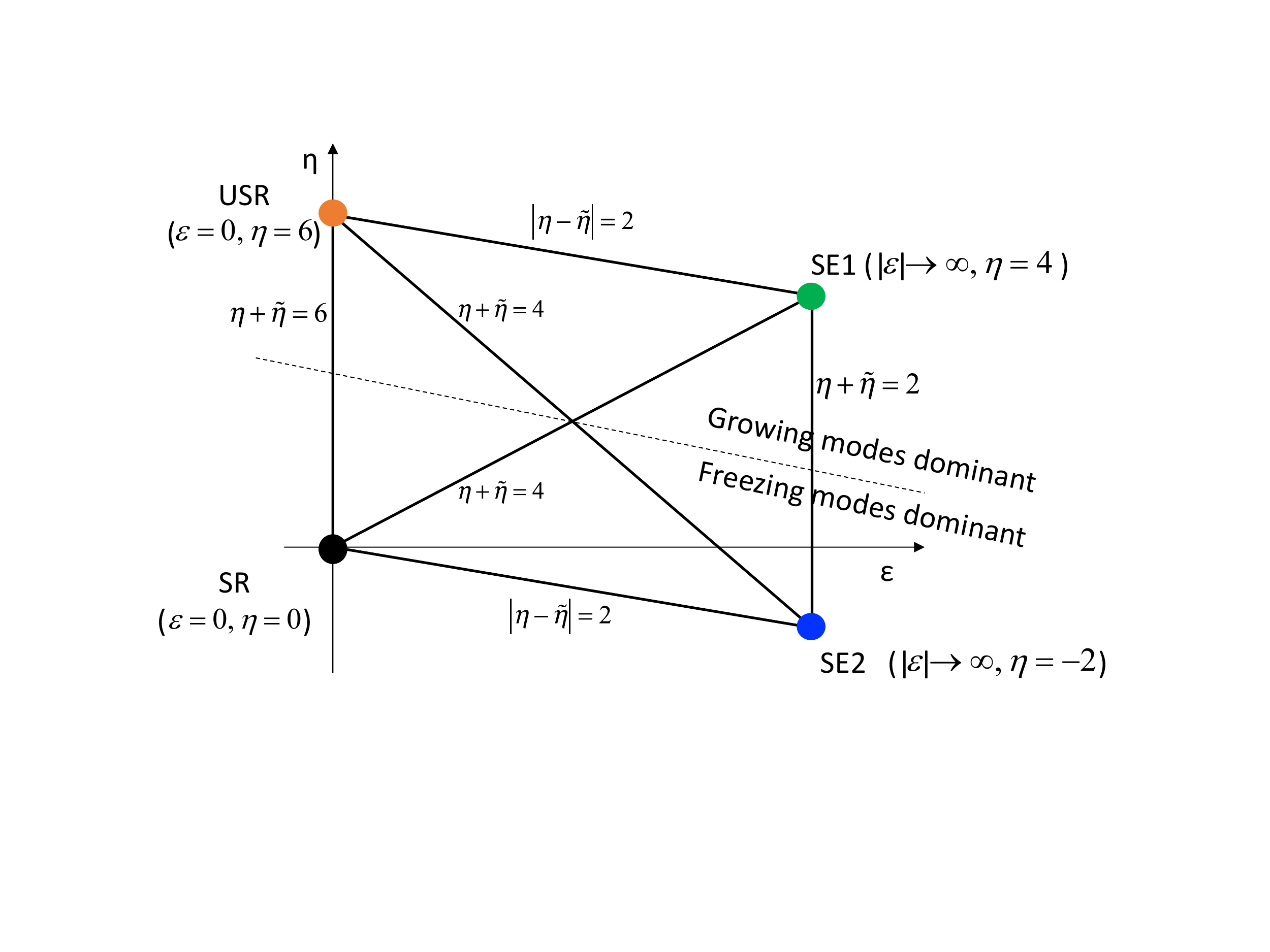}\\
\caption{The sketch plot that demonstrates the dual relationship of models for varying $\epsilon$ and constant $c_s$ in the $(\epsilon, \eta)$-plane. The colored points represent models and the black lines between them  denote various relationships. Moreover, the dashed line divides the whole region into two parts. In the upper parts the perturbations are dominated by the growing modes while in the lower parts the perturbations are dominated by the freezing (constant) modes.   }\label{Sketch1}
\end{figure}

From above one can see that, requiring the scalar spectral index to be identical, we can in total get 4 kind of duality relations of cosmological models with varying slow-roll parameter $\epsilon$. Moreover, taking into account the observational constraint that the scalar spectrum is scale-invariant, we can actually reduce to 4 representative models which, under different relations, are dual to each other: $\epsilon\rightarrow 0$, $\eta\rightarrow 0$ (slow-roll inflation, SR), $|\epsilon|\rightarrow \infty$, $\eta\rightarrow -2$ (slow-evolving universe I, SE1), $|\epsilon|\rightarrow \infty$, $\eta\rightarrow 4$ (slow-evolving universe II, SE2), and $\epsilon\rightarrow 0$, $\eta\rightarrow 6$ (ultra-slow-roll inflation, USR). It is clearer to draw a sketch plot to express these models under the duality relation, as shown in Fig. \ref{Sketch1}.

As a side ramark, we mention that in principle, one can also use Eq. (\ref{nu}) to make up duality relations for models with constant $\epsilon$, such as inflation or MB, just as was done in \cite{Wands:1998yp}.  However, in those cases, the approximations of $(\epsilon-3)/(\epsilon-1)$ will be dependent on specific values of $\epsilon$. Therefore our duality relations will not apply. We will not bring these cases into the current discussion.
\\\\
On the other hand, we can also discuss about the duality relation given by tensor perturbation. According to Eq. (\ref{scalefactor}), it is straightforward to get:
\be
\frac{a^{\prime\prime}}{a}=(1-\epsilon+\eta)^{-2}(-\tau)^{-2}(1+\eta)^2(2-\epsilon)~,
\ee
and from Eq. (\ref{tensorEoM}), one has:
\bea
\nu_T&=&\pm\frac{1}{2}\frac{\epsilon-3(1+\eta)}{\epsilon-(1+\eta)}\sqrt{1-\frac{4\epsilon\eta(\eta+1)}{[\epsilon-3(1+\eta)]^{2}}}~\nonumber\\
&\simeq&\pm\frac{1}{2}\frac{\epsilon-3(1+\eta)}{\epsilon-(1+\eta)}~.
\eea
Similarly, we're using the 0-th approximation for varying $\epsilon$ and when $\eta=0$, the result covers the usual case of $2\nu_T=\pm(\epsilon-3)/(\epsilon-1)$.

One can see that, requiring the duality relation to be maintained also for for tensor spectral index, $|\nu_T|=|\tilde{\nu}_T|$, namely to have
\be
\label{dualtensor}
\frac{\epsilon-3(1+\eta)}{\epsilon-(1+\eta)}=\pm\frac{\tilde{\epsilon}-3(1+\tilde{\eta})}{\tilde{\epsilon}-(1+\tilde{\eta})}~,
\ee
results in that $\epsilon$ and $\tilde{\epsilon}$ must be approaching to the same direction, and $``\pm"$ can only be $``+"$. This means that the duality relations (\ref{dual1}) and (\ref{dual4}) will be broken, while only (\ref{dual2}) and (\ref{dual3}) remains. Therefore, if we detect the tensor spectral index, the dual symmetry among these models will get reduced.

As is well known, the tensor perturbations contribute to the primordial gravitational waves. Note that recently, more and more programs detecting gravitational waves are coming out, among which there are not only those aiming at mediate/low frequency GW (mainly generated by compact binary systems), such as FAST \cite{Nan:2011um}, LISA \cite{AmaroSeoane:2012km}, LIGO \cite{LIGO}, SKA \cite{Janssen:2014dka}, TianQin \cite{Luo:2015ght}, Taiji \cite{Guo:2018npi}, GECAM \cite{Zhang:2018nos}, NANOGrav \cite{Brazier:2019mmu} and so on, but also those aiming at primordial GW program (mainly via polarizations of CMB photons), such as AliCPT \cite{Li:2017lat}, ACT \cite{ACT}, POLARBEAR \cite{POLARBEAR}, SPT \cite{SPT}, BICEP \cite{BICEP}, LiteBIRD \cite{Hazumi:2021yqq} and so on. These programs can make the detections of tensor spectrum (in terms of tensor/scalar ratio $r$), and even tensor spectral index, possible in the future. This will break the duality relation between those models, and thus can differentiate different models of the early universe.



\section{Cosmic duality for both varying $\epsilon$ and $c_s$}

In the following, we will extend our consideration to include the case where $c_s$ is also varying. Assuming that $c_s=c_{s0}(-\tau)^s$, Eq. (\ref{z}) will be modified as:
\bea
\label{z2}
z&\simeq&\sqrt{|\epsilon_{0}|}c_{s0}^{-2}(-\tau)^{\frac{\eta}{2}-1-s}|1+\eta-\epsilon_{0}(-\tau)^{\eta}|^{1/\eta}~, \nonumber\\
\frac{z^{\prime\prime}}{z}&=&\frac{1}{4}(1-\epsilon+\eta)^{-2}(-\tau)^{-2}[(1+\eta)^2(\eta-2-2s)(\eta-4 \nonumber\\
&&-2s)-2\epsilon(\eta+1)(\eta^2-2\eta-4s\eta+8s+4s^2+2) \nonumber\\
&&+\epsilon^2(\eta-2s)(\eta-2-2s)]~,
\eea
and from (\ref{nuz}) we get

\bea
\nu&=&\pm\frac{1}{2(s+1)}\left(\frac{\epsilon-3(1+\eta)}{\epsilon-(1+\eta)}-\eta+2s \right)\nonumber\\
&&\times\sqrt{1-\frac{4\epsilon\eta(\eta+1)}{[\epsilon(1-\eta+2s)+(\eta+1)(\eta-3-2s)]^{2}}}~.\nonumber\\
\eea
In the limit of small $\epsilon$ and large $\epsilon$, we have
\be
\nu=\begin{cases}
\pm\frac{1}{2}\left(2-\frac{\eta+1}{s+1}\right), & \left|\epsilon\right|\gg1,\\
\pm\frac{1}{2}\left(2-\frac{\eta-1}{s+1}\right), & \left|\epsilon\right|\ll1.
\end{cases}
\ee

As shown in the last section, taking into account the tensor spectral index, the two models to be dual must have the same approximate behavior of $\epsilon$, therefore for $\epsilon,\tilde{\epsilon}\rightarrow \pm\infty$, the duality relation for $\nu$ is
\be
\left(2-\frac{\eta+1}{s+1}\right)=\pm\left(2-\frac{\tilde{\eta}+1}{\tilde{s}+1}\right)~.
\ee
When $``\pm"\rightarrow``-"$, the above relation reduces to:
\be
\label{dual2+}
\frac{\eta+1}{s+1}+\frac{\tilde{\eta}+1}{\tilde{s}+1}=4~.
\ee
Note that if we set $s=\tilde{s}=0$, Eq. (\ref{dual2+}) will furtherly reduce to Eq. (\ref{dual2}). In other words, (\ref{dual2+}) will be the generalized version of (\ref{dual2}) by taking into account the varying of sound speed. Moreover, for the case $``\pm"\rightarrow``+"$ which is trivial in the absense of $s,\tilde{s}$, we can also get a somehow nontrivial relation, namely
\be
\frac{\eta+1}{s+1}=\frac{\tilde{\eta}+1}{\tilde{s}+1}~.
\ee
Furthermore, we consider the constraint of scale invariance of the power spectrum, $3-2|\nu|=0$. For $``\pm"\rightarrow``-"$, we have:\\
$\bullet$ $\nu=-\tilde\nu=3/2$, which leads to $\eta=-2-s$, $\tilde\eta=4+5\tilde{s}$. \\

\begin{figure}
\centering
\includegraphics[scale=0.7]{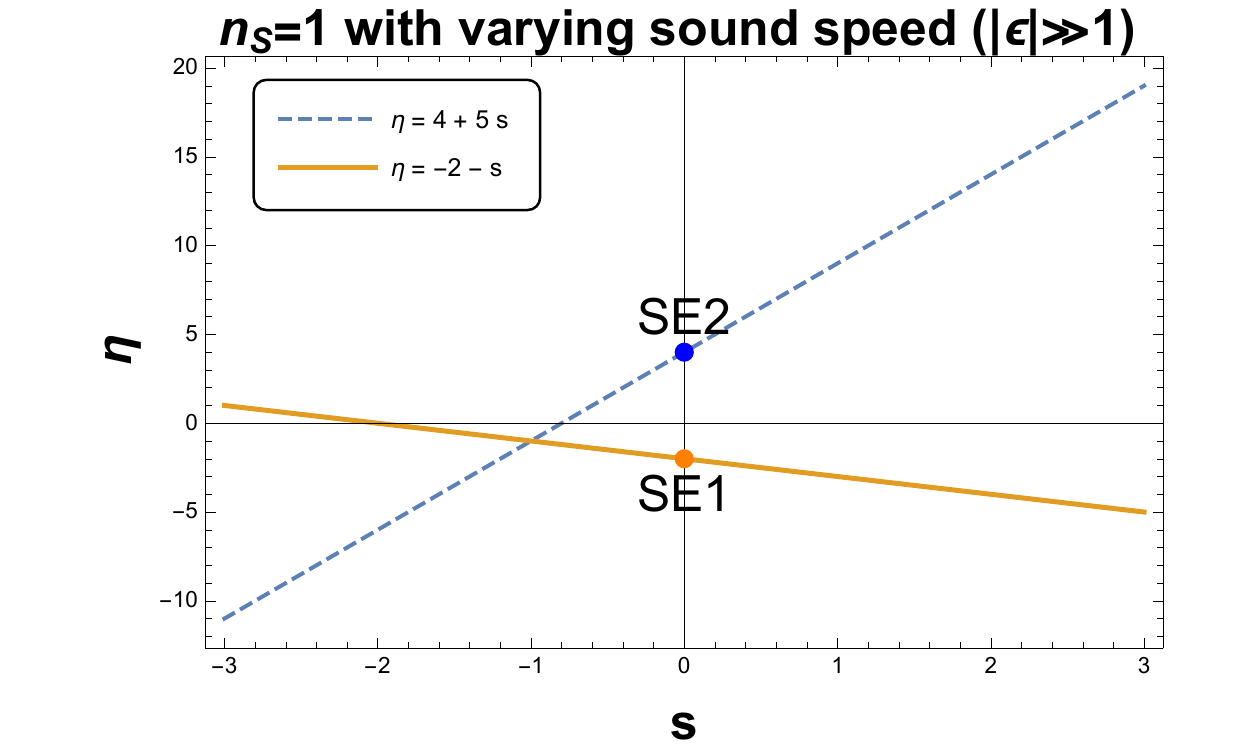}\\
\caption{The sketch plot that demonstrates the dual relationship of models for both varying $\epsilon$ and $c_s$, provided that $\epsilon,\tilde{\epsilon}\rightarrow \pm\infty$. The yellow solid line and the blue dashed line represent models with $\eta=-2-s$ and $\tilde\eta=4+5\tilde{s}$, respectively. The intersection points of the two lines with the vertical axis denotes models with $s=0$ (constant $c_s$), namely the slow-evolving models defined in the context.}\label{Sketch2}
\end{figure}

One could see that the duality between two model-points ($\eta=-2$, $\tilde\eta=-4$) on the $\eta$-axis (1D) in last section has been extended to that of two lines on the $(\eta, s)$-plane (2D). For $\nu=-\tilde\nu$, the two models dual to each other lies on the two lines separately, while for $\nu=\tilde\nu$, as is the case of $``\pm"\rightarrow``+"$, both models will lie on the same line, and which line depends on whether $\nu/\tilde\nu$ is positive or not. Therefore, models presented by either two points lying on those two lines can be dual to each other. In order to illustrate this, we plot the two lines in the $(\eta,s)$-plane in Fig. \ref{Sketch2}. The solid line represents the relation $\eta=-s-2$ while the dashed line represents the another relation $\tilde\eta=4+5\tilde{s}$. We also point out SE1 and SE2 scenarios when $s=0$ by orange point and blue point respectively.

We can also do the same thing for $\epsilon,\tilde{\epsilon}\rightarrow 0$. In this case, the duality relation for $\nu$ is
\be
\left(2-\frac{\eta-1}{s+1}\right)=\pm\left(2-\frac{\tilde{\eta}-1}{\tilde{s}+1}\right)~.
\ee
When $``\pm"\rightarrow``-"$, the above relation reduces to:
\be
\label{dual3+}
\frac{\eta-1}{s+1}+\frac{\tilde{\eta}-1}{\tilde{s}+1}=4~.
\ee
and by setting $s=\tilde{s}=0$, Eq. (\ref{dual3+}) will furtherly reduce to Eq. (\ref{dual3}), namely (\ref{dual3+}) will be the generalized version of (\ref{dual3}) by taking into account the varying of sound speed. Moreover, for the case $``\pm"\rightarrow``+"$ which is trivial in the absense of $s,\tilde{s}$, we can also get a somehow nontrivial relation, namely
\be
\frac{\eta+1}{s+1}=\frac{\tilde{\eta}+1}{\tilde{s}+1}~.
\ee
We also consider the constraint of scale invariance of the power spectrum, $3-2|\nu|=0$. For $``\pm"\rightarrow``-"$, we have:\\
$\bullet$ $\nu=-\tilde\nu=3/2$, which leads to $\eta=-s$, $\tilde\eta=6+5\tilde{s}$. \\

\begin{figure}
\centering
\includegraphics[scale=0.7]{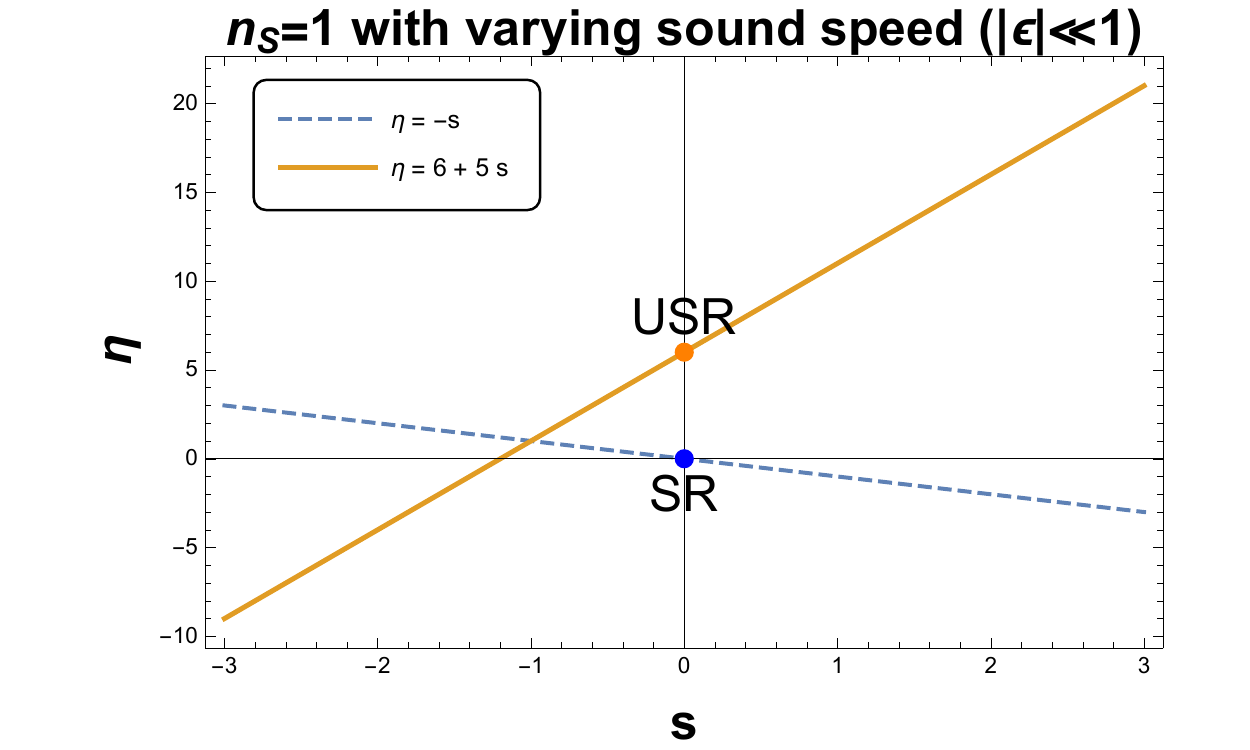}\\
\caption{The sketch plot that demonstrates the dual relationship of models for both varying $\epsilon$ and $c_s$, provided that $\epsilon,\tilde{\epsilon}\rightarrow 0$. The yellow solid line and the blue dashed line represent models with $\eta=6+5s$ and $\tilde\eta=-\tilde{s}$, respectively. The intersection points of the two lines with the vertical axis denotes models with $s=0$ (constant $c_s$), namely the ultra-slow-roll and usual slow-roll inflation models.}\label{Sketch3}
\end{figure}

In like manner as above, models presented by either two points lying on those two lines can be dual to each other. In order to illustrate this, we plot these two lines in Fig. \ref{Sketch3}. The solid line represents the relation $\tilde\eta=6+5\tilde{s}$ while the dashed line represents the another relation $\eta=-s$. The two points where $s=0$ correspond to SR and USR scenarios respectively.

\section{numerical verification}

In the above section, we finished the theoretical analysis of which cosmological models with parametrized slow-roll parameter and sound of speed can give rise to spectral indices that can be dual to each other. The analysis is, however, semi-analytical and several approximations have been used. In order to verify our results, in this section we calculate numerically the equation of motions: Eqs. (\ref{equation}) as well as (\ref{tensorEoM}) with different behaviors of $a(\tau)$, $\epsilon(\tau)$ and $c_s(\tau)$, to see what their tensor and scalar spectra (an their indices) will behave like.

\begin{figure}
  \centering
  \includegraphics[width=7.5cm]{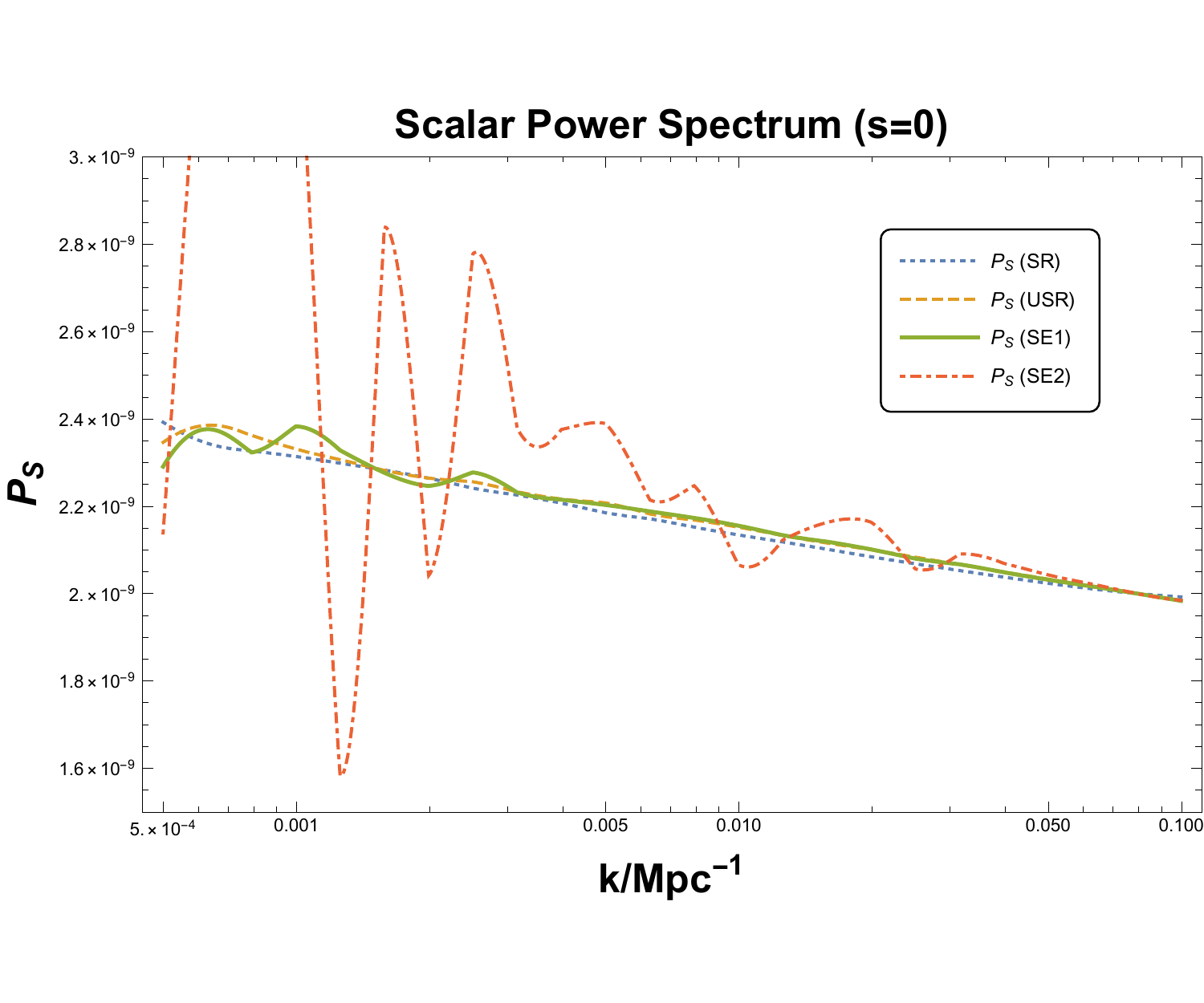}\\
  \caption{The scalar power spectrum with constant sound speed ($s=0$).  The nearly scale-invariant (i.e. $n_S\simeq0.965$ at the pivot scale $k_*=0.05\mathrm{Mpc}^{-1}$) scalar power spectrums against wavenumber $k$ are shown by four lines (blue dotted, orange dashed, green solid and red dot-dashed) corresponding to four different scenarios SR, USR, SE1, SE2, respectively. Here we set $\epsilon<0$, $H>0$ for slow-evolving models, and  $\epsilon>0$, $H>0$  for inflation models.}\label{Fig1}
\end{figure}

\begin{figure}
  \centering
  \includegraphics[width=7.5cm]{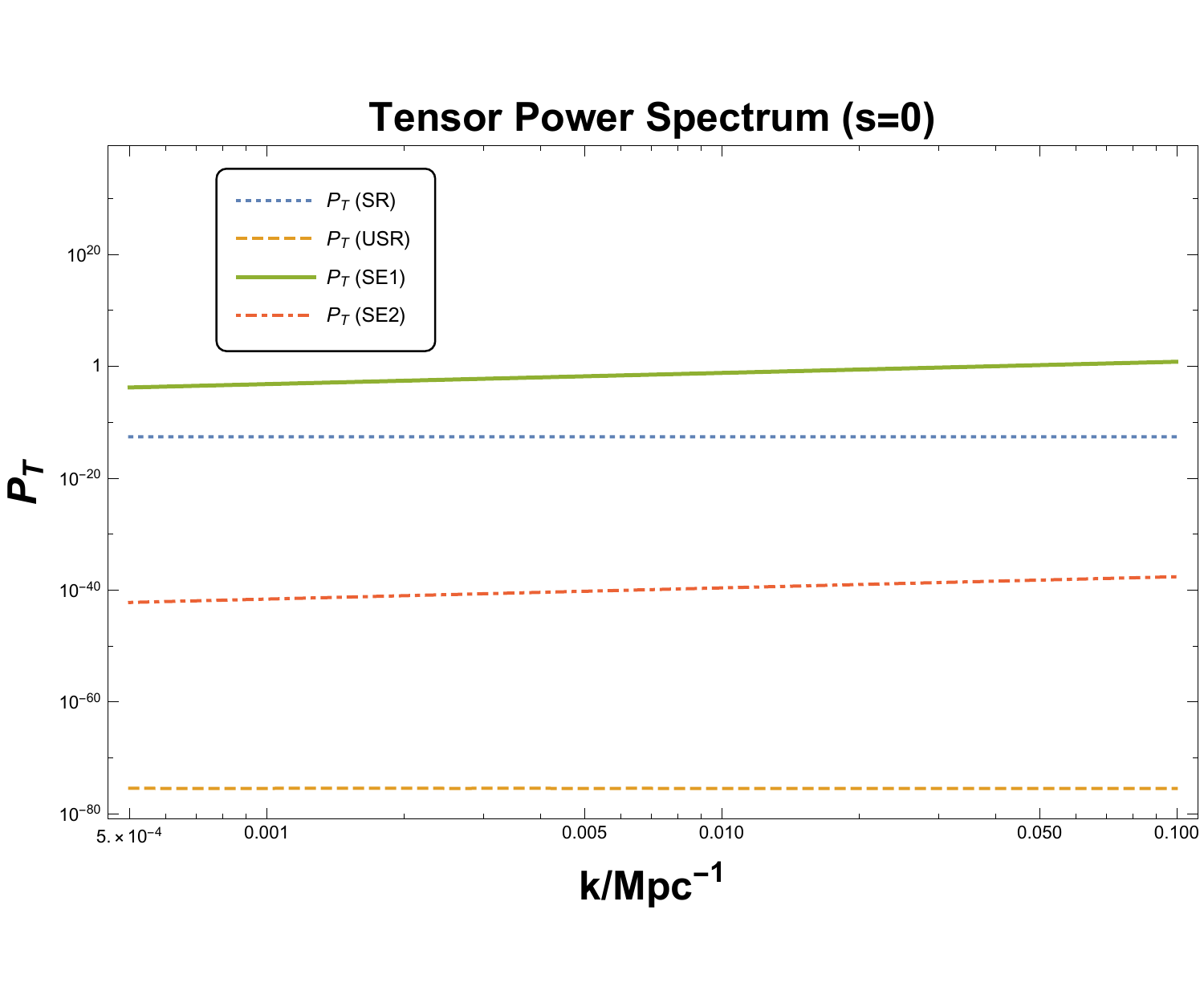}\\
  \caption{The tensor power spectrum with constant sound speed ($s=0$). The tensor power spectrums against wavenumber $k$ are shown by four lines (blue dotted, orange dashed, green solid and red dot-dashed) corresponding to four different scenarios SR, USR, SE1, SE2, respectively. Here we set $\epsilon<0$, $H>0$ for slow-evolving models, and  $\epsilon>0$, $H>0$  for inflation models.}\label{Fig2}
\end{figure}

\begin{figure}
  \centering
  \includegraphics[width=7.5cm]{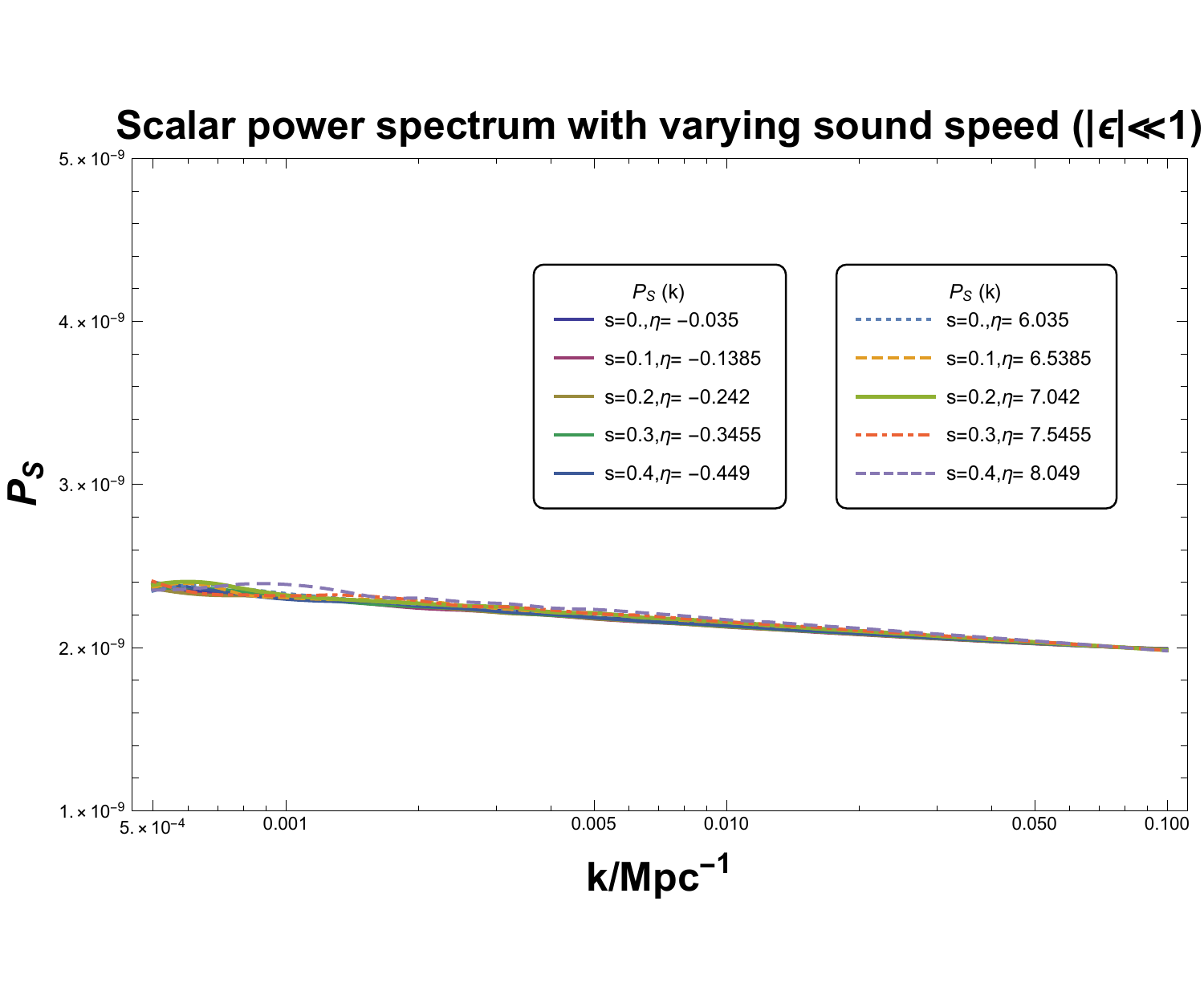}\\
  \includegraphics[width=7.5cm]{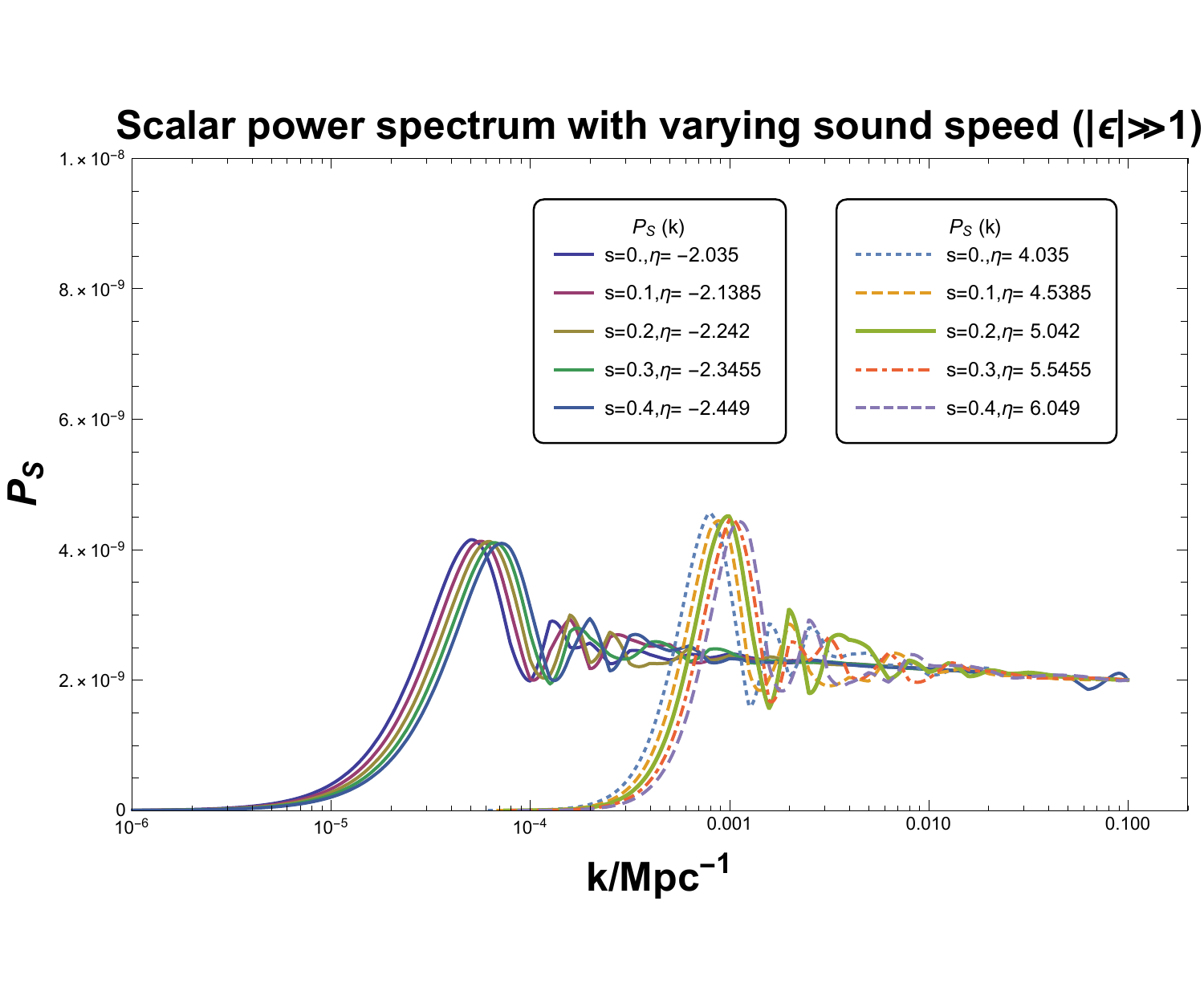}\\
  \caption{The scalar power spectrum with varying sound speed.
   Upper panel: The nearly scale-invariant (i.e. $n_S\simeq0.965$ at the pivot scale $k_*=0.05\mathrm{Mpc}^{-1}$) scalar power spectrums against wavenumber $k$  are shown for SR and USR models ($s=0$) and their variations ($s=0.1,0.2,0.3,0.4$). Lower panel: The nearly scale-invariant scalar power spectrums against wavenumber $k$ (i.e. $n_S\simeq0.965$ at the pivot scale $k_*=0.05\mathrm{Mpc}^{-1}$ ) are shown for SE1 and SE2 models ($s=0$)and their variations ($s=0.1,0.2,0.3,0.4$). Here we set $\epsilon<0$, $H>0$ for slow-evolving models, and  $\epsilon>0$, $H>0$  for inflation models. }\label{Fig3}
\end{figure}

\begin{figure}
  \centering
  \includegraphics[width=7.5cm]{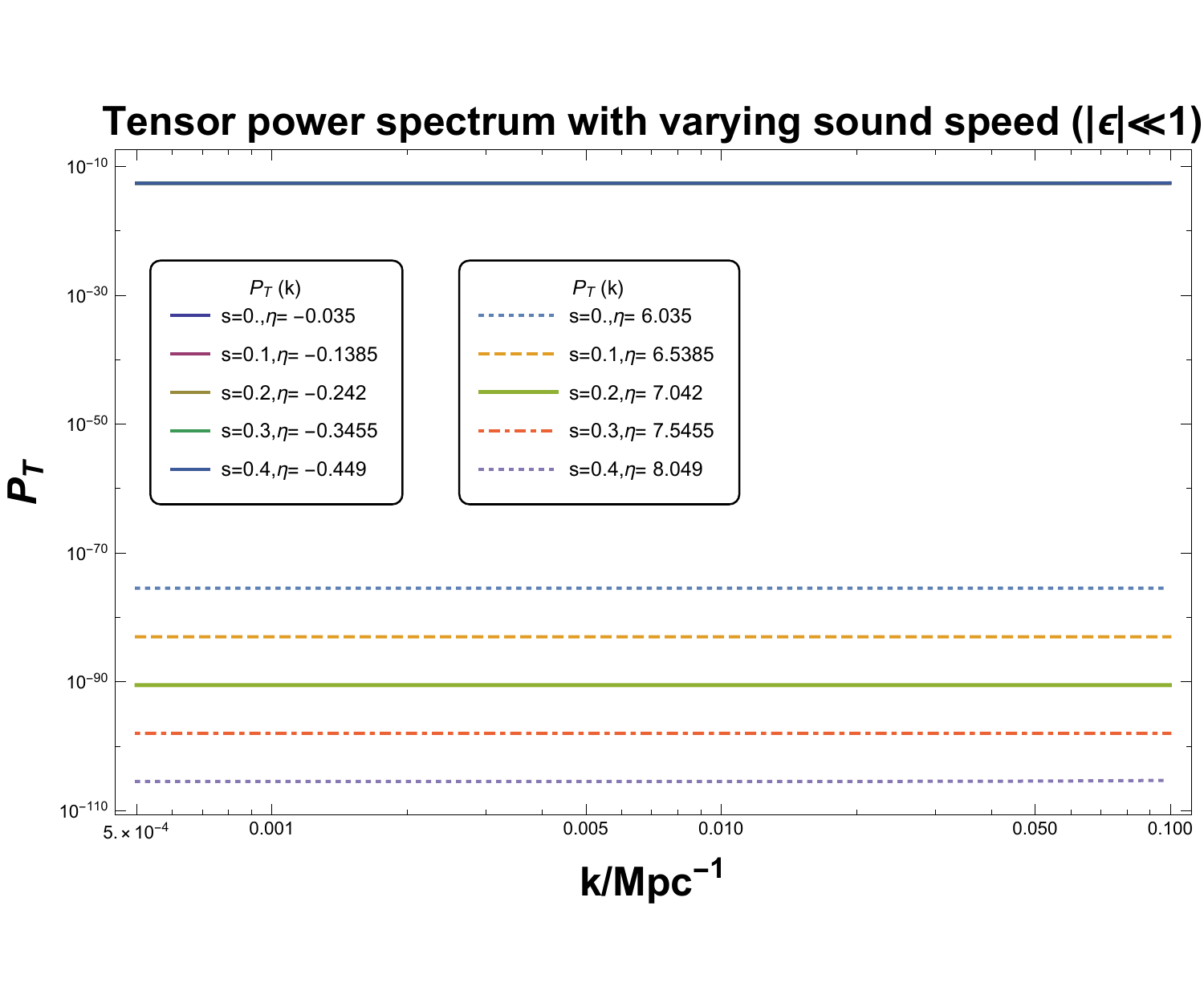}\\
  \includegraphics[width=7.5cm]{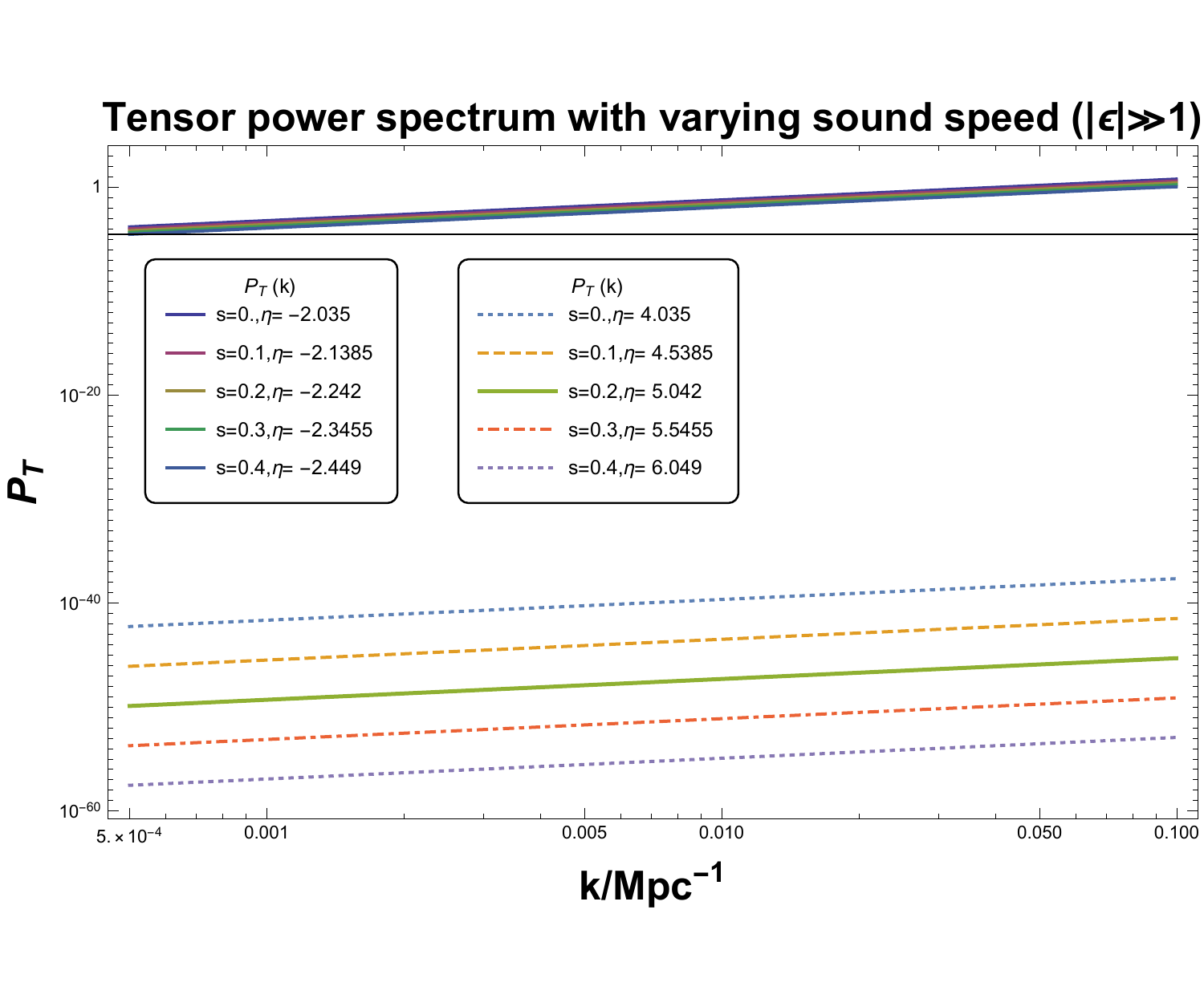}\\
  \caption{The tensor power spectrum with varying sound speed. Upper panel: The tensor power spectrums against wavenumber $k$ are shown for SR and USR models ($s=0$) and their variations ($s=0.1,0.2,0.3,0.4$), where the tensor spectral index $n_T\simeq0$. Lower panel: The tensor power spectrums against wavenumber $k$ are shown for SE1 and SE2 models ($s=0$) and their variations ($s=0.1,0.2,0.3,0.4$), where the tensor spectral index $n_T\simeq2$. Here we set $\epsilon<0$, $H>0$ for slow-evolving models, and  $\epsilon>0$, $H>0$  for inflation models.}\label{Fig4}
\end{figure}

We plot our numerical results for scalar and tensor power spectrum for each typical models in Figs. \ref{Fig1}, \ref{Fig2}, \ref{Fig3} and \ref{Fig4}, respectively. The lines in the figures represent spectrum for each model, while their slopes reflect the information of the spectral indices. One can see from Fig. \ref{Fig1} that, for trivial sound speed case ($s=0$), the spectrum of the four models will have the same behavior at least around the observable range, namely $k\simeq 0.05\text{Mpc}^{-1}$. Moreover, this is not only for indices of the spectra (slope of each line), but for amplitudes as well. The coincidence of the amplitudes can be done by setting proper initial conditions of background quantities such as $a(\tau)$, $H(\tau)$ and $\epsilon(\tau)$. For smaller $k$ region, however, there might be some differences, for example, the slow-evolution models present an oscillating behavior, which might be due to the features in the earlier time that possibly break down some of the approximations in our analytical study. One the other hand, as shown in Fig. \ref{Fig2}, neither amplitude nor slope of the tensor spectra coincide with each other. The reason for slope has already been shown by calculation in the last section, while the reason for amplitude is also understandable: since $P_T/P_S=16|\epsilon|$ and those models have different $\epsilon$, it is impossible to have both $P_S$ and $P_T$ coincide. That means, in $s=0$ case, we can only have scalar power spectrum dual to each other, but not tensor ones.

For $s\neq 0$ case, however, things become different. First of all, as there is one more degree of freedom, the models dual to each other becomes richer. In Fig. \ref{Fig3}, we show that for several choices of $s$, as long as the relationship $\eta=-s$, $\tilde\eta=6+5\tilde{s}$ (upper panel), or $\eta=-s-2$, $\tilde\eta=4+5\tilde{s}$ (lower panel) is satisfied, the amplitude and slope of each line will coincide with each other (Note that in analytical study we approximate the spectral index to be unity but the realistic observation favors $n_S\simeq 0.965$, so the numerical values will be slightly derivated from the analytical formulae). Moreover, for the lower panel for the slow-evolution case, one can see that while the duality happens around the observable range, $k\simeq 0.05\text{Mpc}^{-1}$, it may not be so for smaller $k$ region, due to the same reason as in $s=0$ case.

For tensor spectra, one can see from Fig. \ref{Fig4} that the slopes of each line now get identical, indicating that different from the $s\neq 0$ case, the tensor spectral index can be dual to each other. However, the amplitude of the tensor spectrum still cannot be the same, due to the reason that these models have quite different $\epsilon$ (Although in this case $P_T/P_S=16|\epsilon| c_s$ where $c_s$ can also help to do the modulation, since $c_s$ is contrained to be between $[0,1]$, the modulation is not efficient enough). Therefore with varying sound speed taken into account, only spectral index of scalar and tensor spectrum can be made dual to each other.

The non-duality in the amplitude of tensor perturbations, due to the discrepancy of $\epsilon$, has also been realized in \cite{Gao:2019sbz}, although they have been considering such a problem in the case of scalar perturbations. Can we have the amplitude of tensor spectrum coincide as well, namely, to have full duality of all the quadratic perturbations for cosmological models? Fortunately, the answer maybe ``yes", but some delicate mechanisms may be needed. For example, in \cite{Lin:2019fcz}, the authors suggested that in the case of ultra-slow-roll inflation, the ultra-slow-roll region is not an attractor solution but only a transient phase, which would eventually evolve into the slow-roll phase. Therefore in this model, the perturbations produced will be totally the same as that of the slow-roll inflation models, and there will be fully duality. However, such a mechanism seems model dependent, namely according to each specific model, the details might be different. Since in this paper we're only trying to discuss about the general features without going to details of each model, such mechanisms is somehow beyond the scope of our discussion.

\section{conclusions}
In this paper, we discussed about cosmological models of the early universe, in framework of GR, but relaxing other parameters such as slow-roll parameter and sound speed to be varying quantities. It was found that provided those parameters behave under certain relations, the model will give the same spectral index. Based on the adiabatic mechanism of perturbation production, those relations can be viewed as adiabatic duality relationship that links different models together.

For models in which only slow-roll parameter $\epsilon$ is varying, we found that there are four possible duality relationships between the parameter $\eta$, which is the power-law index of $\epsilon$, or known as the second slow-roll parameter, namely $\eta+\tilde{\eta}=6$, $\eta+\tilde{\eta}=4$, $\eta+\tilde{\eta}=2$ as well as $|\eta-\tilde{\eta}|=2$, depending on the evolution trend of $\epsilon$. However, considering the requirement that the scalar power spectrum must be nearly scale-invariant, we found that only four kinds of models as well as matter bounce model that could dual to each other. Moreover, when tensor power spectrum and spectral index is taken into account, the duality relation will be broken.

We also extended the discussion to wider case, where the sound speed $c_s$ is varying as well. In this case, there are two duality relationships, namely $\eta=-s$, $\tilde\eta=5\tilde{s}+6$ and $\eta=-s-2$, $\tilde\eta=5\tilde{s}+4$, but the models dual to each other get enlarged, even scale-invariance of scalar perturbation is still required. Moreover, the spectral index of tensor spectrum can also be dual, although the amplitude of the tensor spectrum cannot. Therefore in contrast to conformal duality, the adiabatic duality might not be a full duality of early time cosmological models. Although all the models can be made within the current bound of tensor perturbations, the future GW detectors may have the power of differentiate these models on the observational side.

We checked all the above results by performing numerical calculations. We also mentioned that, via some specific mechanisms, we may have chance to have full duality among cosmological models. However, these mechanisms seems to be model-dependent and have to be studied case by case. We will leave the related discussions for a separate work.

\begin{acknowledgments}
This work was supported by the National Natural Science Foundation of China under Grants No.~11653002 and No.~11875141.  J.S. was partially supported by the Fundamental Research Funds for the Central Universities (Innovation Funded Projects) under Grants No.~2020CXZZ105.

\end{acknowledgments}

\end{document}